\documentclass[twocolumn,showpacs,prl,english,showkeys,aps,unsortedaddress,superscriptaddress,nofootinbib]{revtex4-2}
\usepackage{amsmath}
\usepackage{amssymb}
\usepackage{epsfig}
\usepackage{graphicx}
\usepackage[T1]{fontenc}
\usepackage[utf8]{inputenc}
\usepackage{color}
\usepackage[normalem]{ulem}
\usepackage{tikz}
\usetikzlibrary{positioning}

\usepackage[colorlinks,allcolors=blue]{hyperref}

\begin{document}

\title{Dynamically Reentrant Skyrmion Phase in Oscillating Magnetic Fields}
\author{Z. Demir Vatansever}
\email{zeynep.demir@deu.edu.tr}
\affiliation{Department of Physics, Dokuz Eyl\"{u}l
University, TR-35160, Izmir-Turkey}
\author{E. Vatansever}
\email{erol.vatansever@deu.edu.tr}
\affiliation{Department of Physics, Dokuz Eyl\"{u}l
University, TR-35160, Izmir-Turkey}
\author{A. Vasilopoulos}
\email{alex.vasilopoulos@essex.ac.uk}
\affiliation{School of Mathematics, Statistics and Actuarial Science, University of Essex, Colchester CO4 3SQ, United Kingdom}
\author{N. G. Fytas}
\email{nikolaos.fytas@essex.ac.uk}
\affiliation{School of Mathematics, Statistics and Actuarial Science, University of Essex, Colchester CO4 3SQ, United Kingdom}
\author{A. Berger}
\email{a.berger@nanogune.eu}
\affiliation{CIC nanoGUNE BRTA, E-20018 Donostia-San Sebastian, Spain}

\date{\today}

\begin{abstract}
We address a central open question in driven topological matter by investigating the nonequilibrium response of a two-dimensional magnet with Dzyaloshinskii–Moriya interactions subjected to a periodically oscillating magnetic field. Using extensive Monte Carlo simulations, we uncover a dynamically re-entrant skyrmion phase that emerges within a finite window of field amplitudes and bias fields, sandwiched between the conventional skyrmion and polarized ferromagnetic phases. This phase has no equilibrium counterpart and originates from an imbalance between skyrmion creation and annihilation over a single field cycle, leading to a reversal of the period-averaged topological charge. We show that the re-entrant phase exists only in the dynamic response regime where the system can follow the drive, and is suppressed at higher driving frequencies. Our results establish periodic driving as a nonequilibrium control parameter for stabilizing and reversing skyrmion topology.
\end{abstract}

\maketitle

Magnetic skyrmions are topologically nontrivial, whirl-like spin textures that have emerged as promising information carriers for spintronic applications owing to their nanoscale size and efficient current-driven dynamics~\cite{Nagaosa2013, Goebel2021}. Beyond their technological relevance, these topological objects provide a fertile platform for exploring novel physical phenomena including robust stabilities protected against external perturbations~\cite{Fert2013}. Originally proposed on theoretical grounds~\cite{Rossler2006}, magnetic skyrmions were later observed in bulk noncentrosymmetric magnets~\cite{Muhlbauer2009, Yu2010, Munzer2010, Yu2010FeGe} and in thin-film systems with broken interfacial inversion symmetry~\cite{MoreauLuchaire2016, Soumyanarayanan2017, Legrand2020}. Depending on the underlying microscopic interactions, skyrmions may originate from distinct mechanisms, including Dzyaloshinskii--Moriya interactions (DMIs)~\cite{Liang2020Janus, Yuan2020, Iroulart2024} and magnetic frustration~\cite{Okubo2012, Leonov2015, Hayami2016, Amoroso2020}. Besides ordered skyrmion lattices, experiments and simulations have revealed the existence of disordered skyrmion phases, often referred to as skyrmion liquids, which lack long-range crystalline order~\cite{Nishikawa2019, Huang2020, Mitsumoto2022}.

In magnetic systems lacking inversion symmetry, spin--orbit coupling gives rise to an antisymmetric Dzyaloshinskii-Moriya type exchange interaction that favors chiral spin textures and stabilizes isolated skyrmions and skyrmion lattices through its competition with Heisenberg type exchange and Zeeman interactions. Hereby, skyrmions are characterized by a topological charge
\begin{equation}\label{top_charge}
\chi_Q=\frac{1}{4\pi}\int d x\, d y \;
\mathbf{S} \cdot
\left( \partial_x \mathbf{S} \times \partial_y \mathbf{S} \right)
\end{equation}
which quantifies the topological nature of the spin configuration. In equilibrium, sufficiently strong magnetic fields destabilize skyrmions in favor of a uniformly polarized ferromagnetic state, thereby confining their stability to a finite field window between low-field magnetically ordered phases (e.g., helical or stripe-domain states) and the high-field ferromagnetic phase. However, this equilibrium picture can be profoundly modified by time-dependent magnetic fields, which open new pathways for topological transformations.

Dynamically driven nonequilibrium phase transitions arise in a wide variety of interacting many-particle systems subjected to oscillatory control parameters~\cite{Henkel, Rao}. A paradigmatic example is provided by dynamic phase transitions in ferromagnets driven by periodically oscillating magnetic fields~\cite{Chakrabarti, Yuksel}. First observed in the kinetic Ising model~\cite{Tome}, dynamic phase transitions manifest through qualitative changes in the magnetization when the field period is varied below the Curie temperature~\cite{Riego0}. Subsequent studies have explored these transitions using mean-field approaches~\cite{Zimmer, Buendia1, Jang1, Shi, Punya, Gallardo, Riego}, Monte Carlo simulations~\cite{Korniss, Lo, Acharyya1, Acharyya2, Acharyya3, Buendia2}, and experiments~\cite{Le, Jiang, Suen, Kleemann, Robb0, Berger, Riego2017, Quintana0}. A remarkable outcome is the close correspondence between equilibrium and dynamic critical behavior, including identical universality classes revealed by finite-size scaling analyses~\cite{Sides1, Sides2, Korniss, Buendia3, Vatansever2, Park2, Vatansever_Fytas, Vatansever3, Vasilopoulos, Vatansever4} and confirmed experimentally in ultrathin Co films~\cite{Quintana0}.  Dynamical studies of magnetic skyrmions have mainly addressed the stability of individual skyrmions~\cite{Hagemeister2015, Bessarab2015, Rozsa2016}, whereas the collective nonequilibrium dynamics of extended skyrmion assemblies remain far less explored. Moreover, collective skyrmion dynamics can be induced by various external stimuli, such as applied fields~\cite{Zhang2022, Dai2023} or current injection~\cite{Yu2012, Sampaio2013, Woo2016}. While device-oriented studies naturally emphasize electrically driven mechanisms, time-dependent magnetic fields provide a general framework for probing collective dynamical phenomena.

In this Letter, we study the nonequilibrium dynamics of a two-dimensional magnet with DMIs under a periodically oscillating magnetic field. We consider a classical two-dimensional ferromagnetic Heisenberg model on a triangular lattice, described by the Hamiltonian
%\begin{widetext}
\begin{equation}
\begin{aligned}
\mathcal{H} =
&- J \sum_{\langle i,j \rangle} \mathbf{S}_i \cdot \mathbf{S}_j
- \sum_{\langle i,j \rangle} \mathbf{D}_{ij} \cdot
\left( \mathbf{S}_i \times \mathbf{S}_j \right) \\
&- A \sum_{i} (S_i^z)^2
- H(t) \sum_{i} S_i^z .
\end{aligned}
\end{equation}
%\end{widetext}
\noindent Here $\mathbf{S}_i$ is a classical unit vector at lattice site $i$. The first term represents the nearest-neighbor Heisenberg-type ferromagnetic exchange interaction with coupling constant $J$. The second term describes the DMI, with
\begin{equation}
\mathbf{D}_{ij} = D \, \frac{\mathbf{r}_i - \mathbf{r}_j}{|\mathbf{r}_i - \mathbf{r}_j|} \times \hat{z},
\end{equation}
where $\mathbf{r}_i$ and $\mathbf{r}_j$ denote the real-space position vectors of spins at lattice sites $i$ and $j$, respectively, and for which the interaction strength is fixed to $D=1$ in our study here. In the present work, we focus on the DMI symmetry that stabilizes Néel-type skyrmions, yet we have also confirmed in Fig.~S3 of~\cite{supplemental} that qualitatively similar results are obtained for Bloch-type skyrmions~\cite{Kezsmarki2015}. The third term denotes an easy-axis single-ion anisotropy along the out-of-plane ($z$) direction, with $A = 0.1$. The last term accounts for the coupling to an external magnetic field, $H(t)$, applied along the $z$ axis, consisting of a static bias field $h_b$ and a time-dependent component $h(t)$, chosen as a square wave with amplitude $h_0$ and half-period $t_{1/2}$. All sums $\langle i,j \rangle$ run over nearest-neighbor pairs. For simplicity, we set $k_{\rm B} = 1$ and $J = 1$~\cite{ParExp1}.

\begin{figure}[h!]
   \includegraphics[scale=0.5]{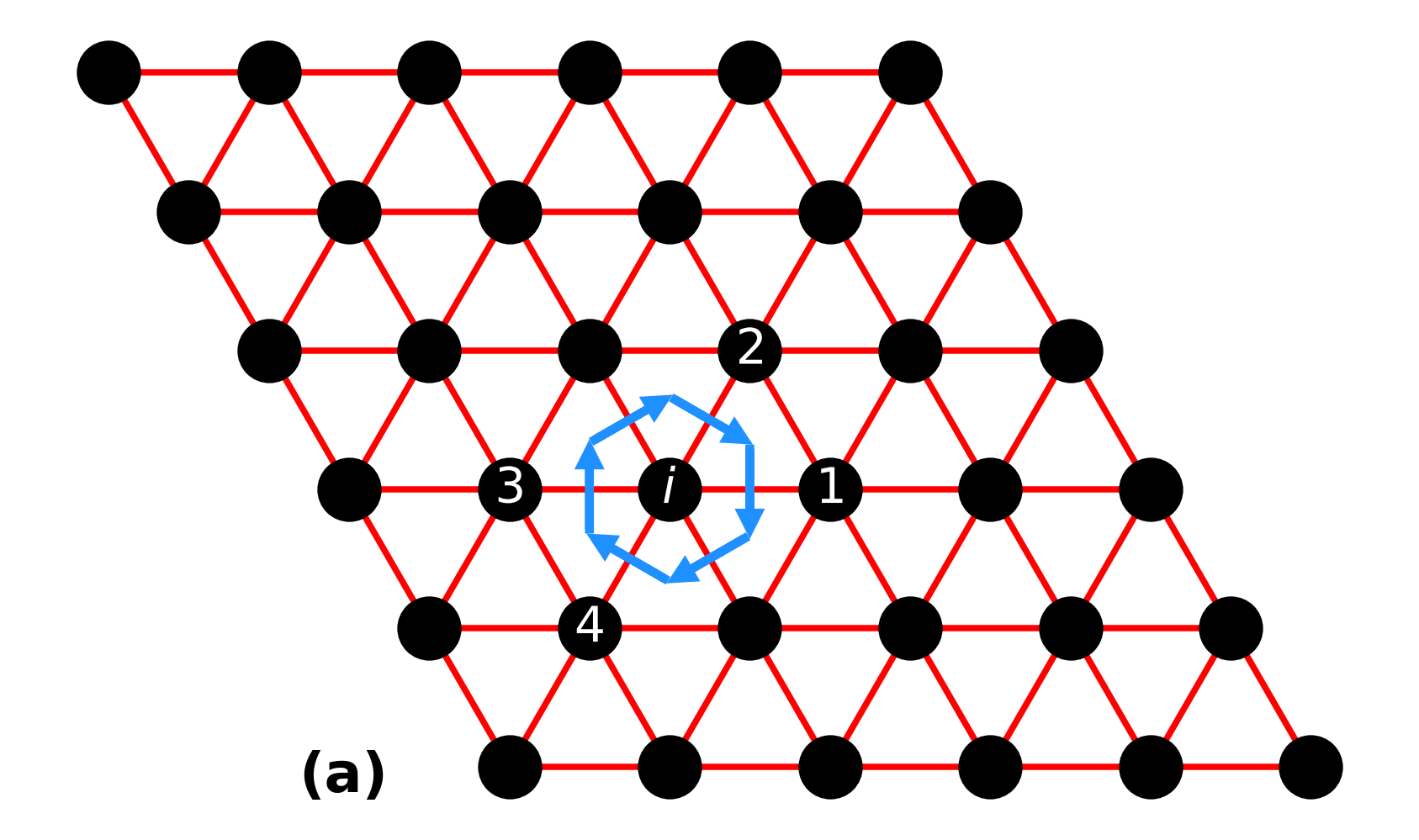}
    \includegraphics[scale=0.5]{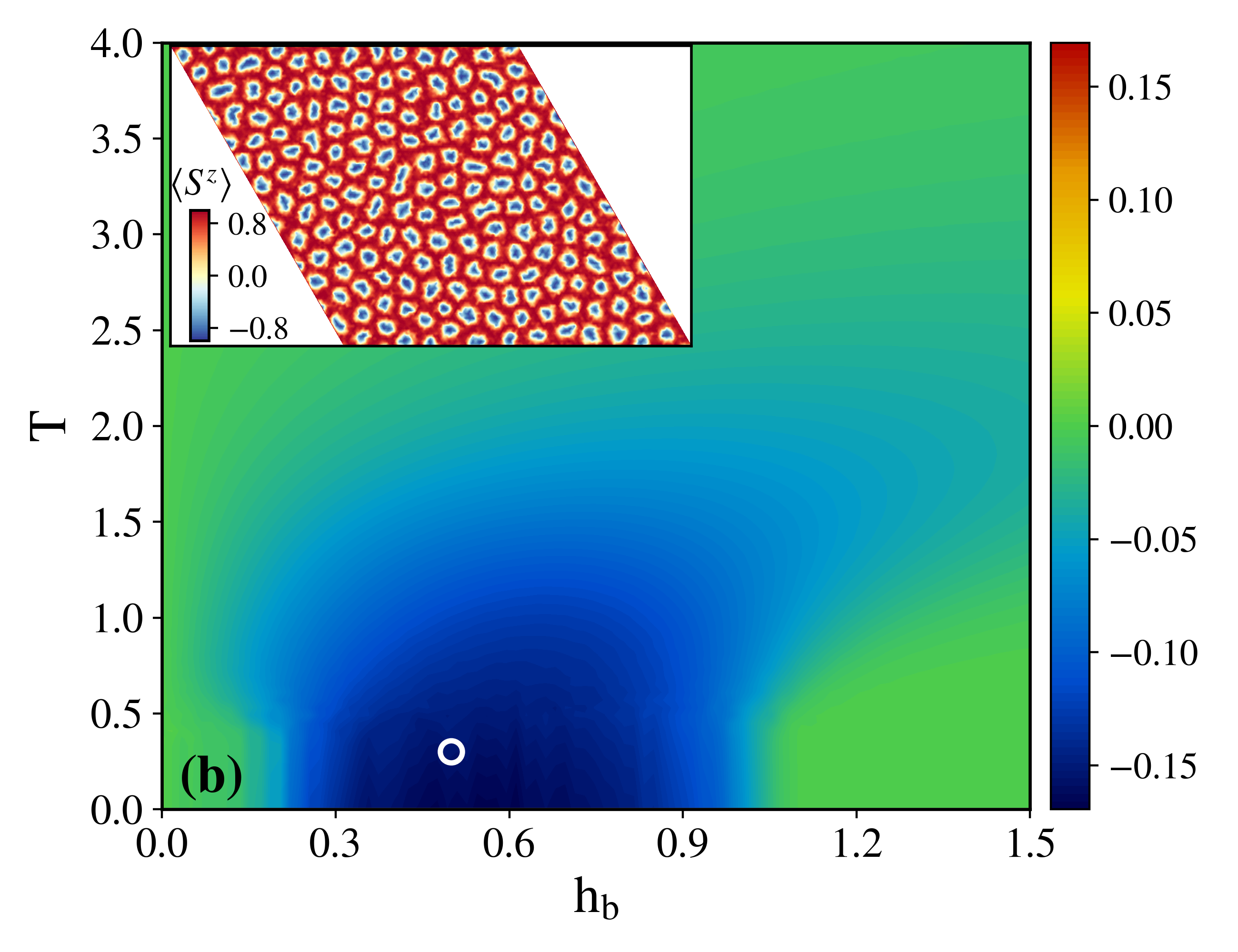}
\caption{(a) Schematic illustration of the triangular lattice geometry used in this work. Blue arrows indicate the DMI vectors $\mathbf{D}_{ij}$ between nearest-neighbor spins for positive $D$. 
(b) Equilibrium skyrmion number as a function of the magnetic field $h_b$ and temperature $T$. The inset shows representative real-space spin textures at the white point marked in the phase diagram. The color scale denotes the out-of-plane spin component $S^z$.
}\label{Fig1}
\end{figure}

Monte Carlo simulations based on the single-spin Metropolis algorithm~\cite{Metropolis,Binder,Newman,Binder2} were performed on triangular lattices of linear size $L=120$ with periodic boundary conditions, as schematically illustrated in Fig.~\ref{Fig1}(a). Finite-size effects were assessed by additional simulations at larger system sizes, which produced essentially identical results. The system is gradually cooled from high temperatures, and after an initial thermalization stage at each temperature, simulations are carried out over $10^{3}$ periods of the external field, during which thermal averages of physical observables are accumulated. Time is measured in units of one Monte Carlo step per site (MCSS). Further algorithmic details are provided in~\cite{supplemental}. 

In periodically driven spin systems, the period-averaged magnetization $Q$ is commonly employed as a dynamical order parameter and is defined as~\cite{Tome}
\begin{equation}
Q = \frac{1}{P} \int_0^{P} M(t)\, dt,
\end{equation}
where $M(t)$ is the instantaneous magnetization per spin, obtained from the sum of the out-of-plane spin components $S_i^z$, and $P=2t_{1/2}$ denotes the period of the oscillating field. 
To characterize the topology, we compute the skyrmion number (topological charge) $\chi_Q$, defined in Eq.~(\ref{top_charge}). In the simulations, $\chi_Q$ is evaluated using a lattice discretization on elementary triangles (see the detailed discussion in~\cite{supplemental}). Under a time-dependent magnetic field, $\chi_Q(t)$ becomes a time-dependent quantity. 
We therefore introduce the period-averaged skyrmion number,
\begin{equation}
\langle \chi_Q \rangle = \frac{1}{P} \int_0^{P} \chi_Q(t)\, dt,
\end{equation}
which serves as a dynamical observable of the topological response of the system. We point out here that within our convention a reversal of the magnetic field changes the sign of the topological charge without implying the formation of antiskyrmions. Throughout this work, $\langle \chi_Q \rangle$ is reported per $10^{3}$ unit cells.

The equilibrium topological response, quantified by the skyrmion number $\chi_Q$, in the $(h_b,T)$ plane is shown in Fig.~\ref{Fig1}(b). 
\begin{figure}[h!]
\includegraphics[scale=0.5]{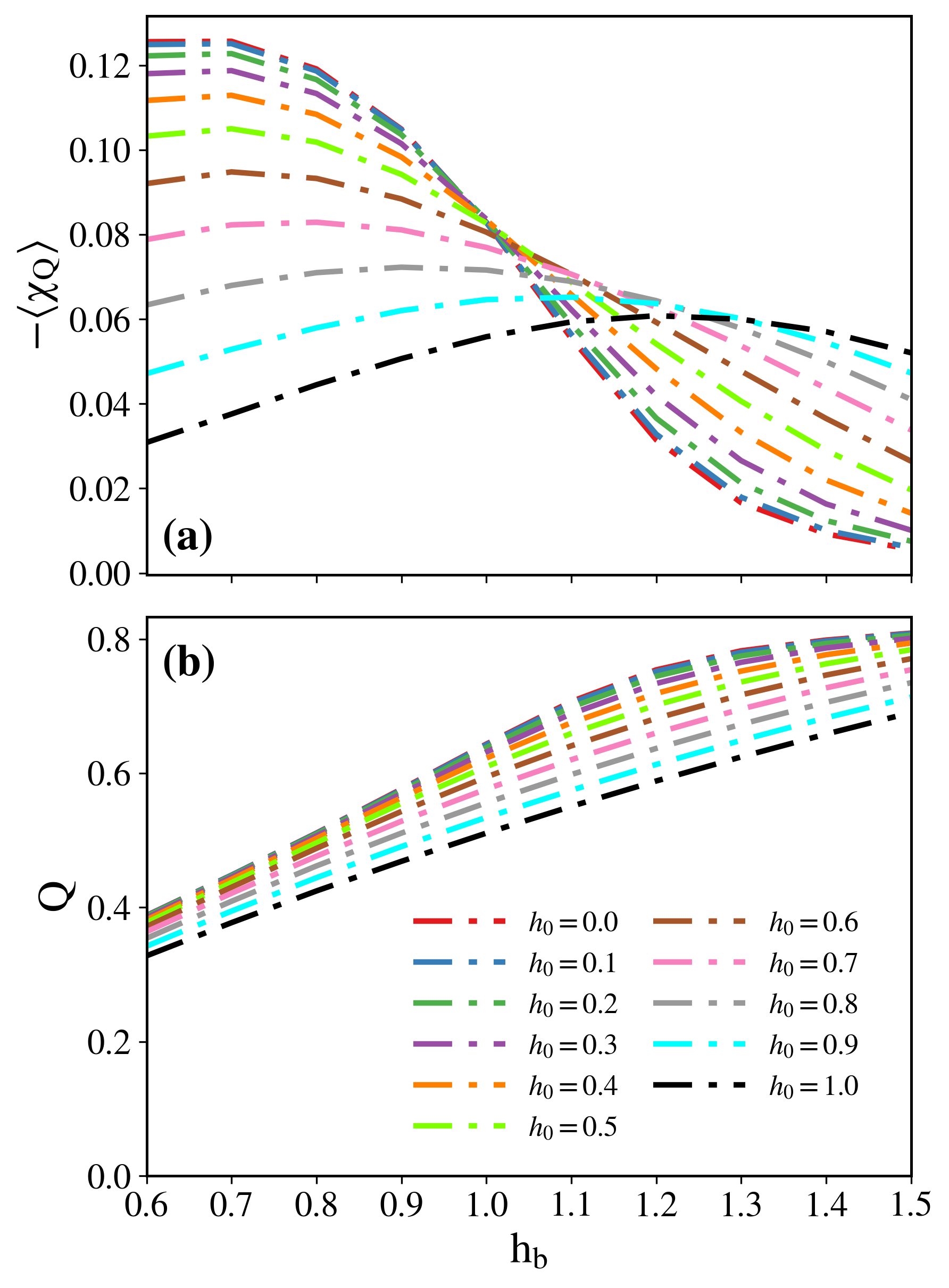}
\caption{(a) Period-averaged skyrmion number $\langle \chi_Q \rangle$ and (b) period-averaged magnetization $Q$ as a function of the bias field $h_b$ for several values of the oscillating field amplitude $h_0$, calculated for a half-period $t_{1/2}=100$ MCSS.}\label{Fig2}
\end{figure}
A blue region with negative $\chi_Q$, corresponding to the skyrmion region, appears at intermediate bias fields and low-to-moderate temperatures due to the competition between Heisenberg exchange, DMIs, easy-axis anisotropy, and Zeeman energy.
The inset displays a representative real-space spin texture from this region, specifically the one at the phase space point identified by the white circle in Fig.~\ref{Fig1}(b), forming a dense skyrmion pattern consistent with the negative topological charge.
At low bias fields, $\chi_Q$ is negligible and stripe-domain configurations dominate. With increasing $h_b$, skyrmions emerge over a finite field window, beyond which the system crosses over to a uniformly polarized ferromagnetic phase with $\chi_Q \simeq 0$. As the temperature rises, the skyrmion region shrinks and eventually vanishes, as thermal fluctuations destroy the topological order and drive the system into a paramagnetic state. A similar behavior has also been reported in realistic material-based models~\cite{Liang2020Janus}.

Motivated by the equilibrium behavior in Fig.~\ref{Fig1}(b), we next investigate the nonequilibrium response of the system under a time-dependent magnetic field $h(t)$. Unless stated otherwise, results are shown at $T=1$, a representative point within the equilibrium skyrmion region. 
\begin{figure}[h!]
\begin{center}
\includegraphics[scale=0.5]{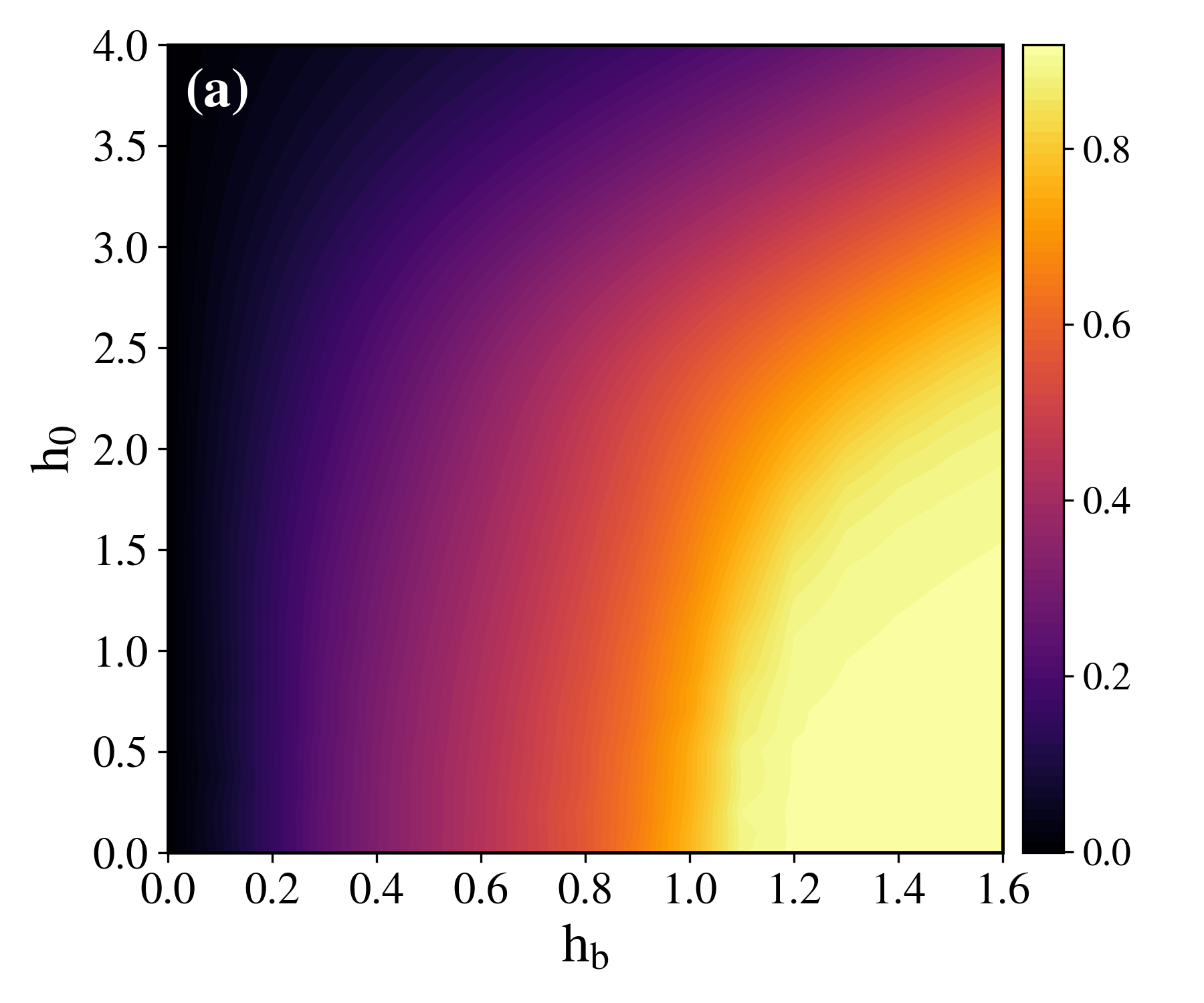}
\includegraphics[scale=0.5]{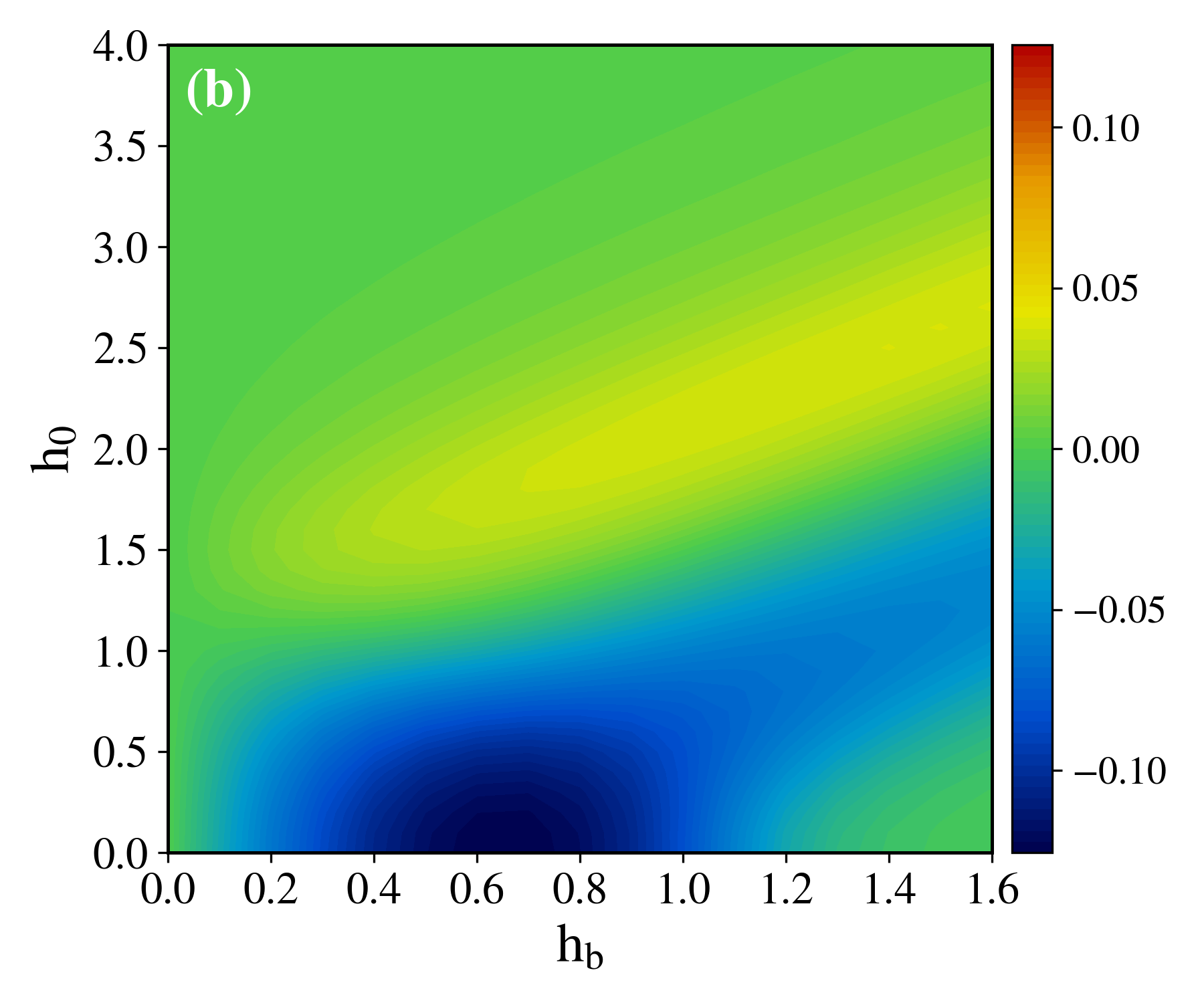}
\caption{(a) Period-averaged magnetization $Q$ and 
(b) period-averaged skyrmion number $\langle \chi_Q \rangle$ in the $(h_b,h_0)$ plane, computed at $t_{1/2}=100$ MCSS. The color scale indicates the magnitude of the corresponding quantity in each panel.}
\label{Fig3}
\end{center}
\end{figure}
Figure~\ref{Fig2}(a) displays the period-averaged skyrmion number $\langle \chi_Q \rangle$ as a function of the bias field $h_b$ for different oscillating-field amplitudes at half-period $t_{1/2}=100$ MCSS, corresponding to a dynamic response regime in which the system can accommodate the driving field. The $h_b$ range is chosen based on the equilibrium phase diagram to span the crossover from the skyrmion phase to the polarized ferromagnetic phase. 
In the absence of an oscillating field ($h_0=0$), $\langle \chi_Q \rangle$ decreases smoothly with increasing $h_b$, consistent with the equilibrium crossover to the polarized phase with vanishing $\langle \chi_Q \rangle$. 
Introducing a time-dependent field suppresses the period-averaged skyrmion number within the skyrmion regime by reducing the stability of skyrmion textures through the time dependence of the Zeeman term. For small $h_0$, $\langle \chi_Q \rangle$ retains a bias-field dependence similar to the static case, while for larger $h_0$, the transition broadens very relevantly.

To compare the evolution of topological order with conventional magnetic order, we analyze the period-averaged magnetization $Q$. As shown in Fig.~\ref{Fig2}(b), $Q$ increases monotonically with $h_b$ for all $h_0$, while the approach to saturation slows progressively with increasing oscillating-field amplitude. At low bias fields, $Q$ remains small due to the DMI, which favors noncollinear spin textures and suppresses ferromagnetic alignment.

The bias-field dependence of the period-averaged skyrmion number already indicates that the topological response varies significantly with the oscillation amplitude. To obtain a global perspective, we explore the full $(h_b,h_0)$ parameter space. Figure~\ref{Fig3} presents the period-averaged magnetization and skyrmion number for a half-period $t_{1/2}=100$ MCSS. As shown in Fig.~\ref{Fig3}(a), $Q$ increases monotonically with $h_b$ for all oscillation amplitudes $h_0$, consistent with the behavior observed in Fig.~\ref{Fig2}(b). Increasing $h_b$ drives the system toward a polarized state, while the oscillating magnetic field shifts the onset of saturation to higher bias fields without changing the overall monotonic trend. 
Although the conventional order parameter behaves as expected, by exhibiting a simple transition in between two equilibrium states that is being broadened and shifted by the oscillating magnetic field, the period-averaged skyrmion number, seen in Fig.~\ref{Fig3}(b), reveals a nontrivial nonequilibrium behavior in the $(h_b,h_0)$ plane. For small field amplitudes ($h_0 \lesssim 1$), a negative region (blue) appears between the stripe-domain and polarized phases, so that this region closely resembles the equilibrium behavior of Fig.~\ref{Fig1}(b), indicating that weak oscillations minimally perturb the system. At intermediate amplitudes and bias fields $h_b \gtrsim 0.4$, however, $\langle \chi_Q \rangle$ re-emerges with the opposite topological sign, forming the yellow region in the phase diagram. This suggests that the skyrmion number behavior does not resemble the simple transition from one limiting behavior to the opposite, shown by $Q$, but instead displays the occurrence of a non-trivial intermediate state. 
\begin{figure}[h!]
\begin{center}
\includegraphics[scale=0.5]{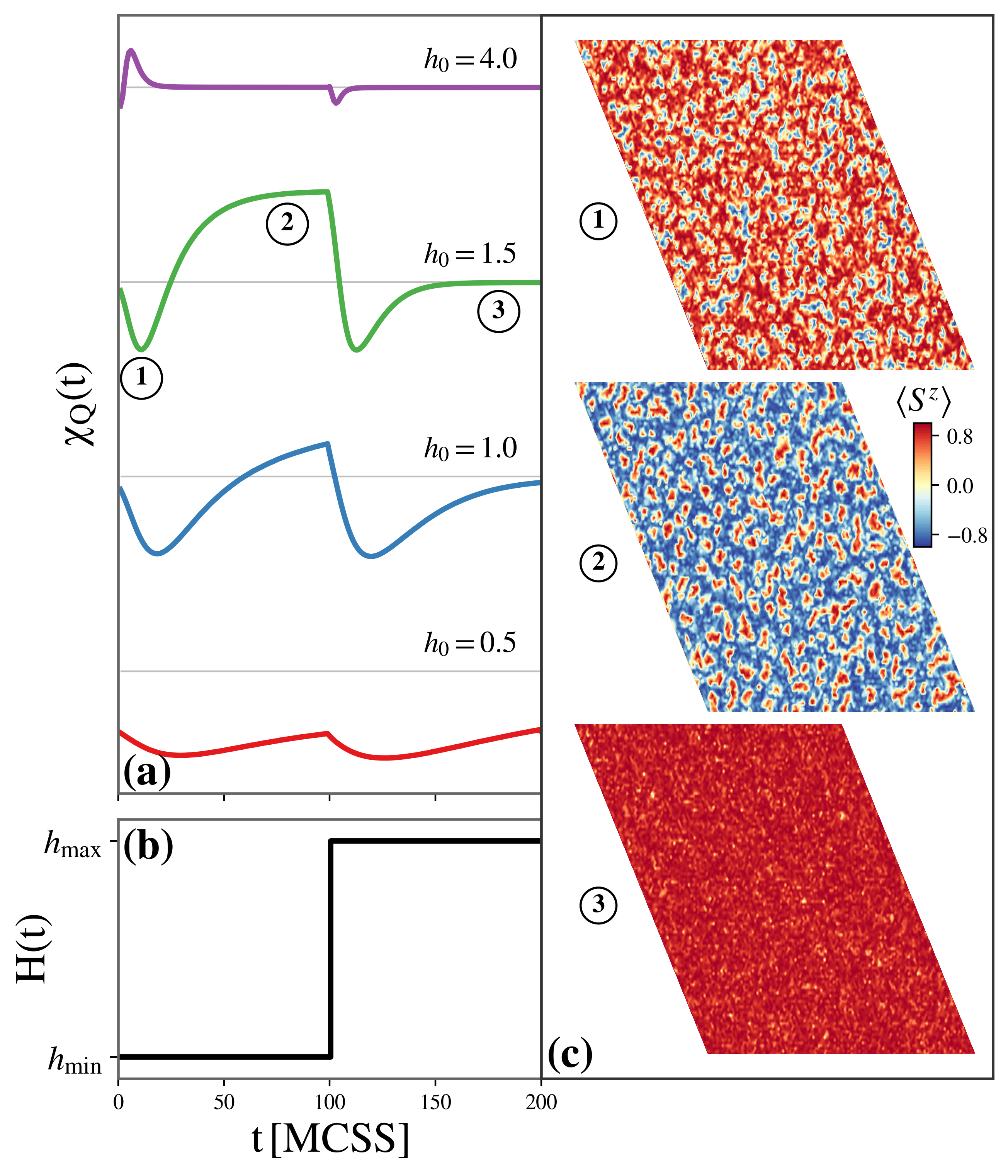}
\caption{Time evolution of: (a) the skyrmion number $\chi_Q(t)$ for several values of $h_0$ with $h_b=0.7$, and (b) the total magnetic field over one cycle of the oscillating field. (c) Real-space spin textures corresponding to the circled points 1, 2, and 3 in panel (a) for $h_0 = 1.5$. The color scale denotes the out-of-plane spin component $S^z$. The half-period is fixed at $t_{1/2}=100$ MCSS. The gray horizontal lines in panel (a) represent $\chi_Q = 0$.} \label{Fig4}
\end{center}
\end{figure}
The observed re-emergence of $\langle \chi_Q \rangle$ with opposite sign defines a dynamically re-entrant topological response, in which skyrmions undergo an imbalanced creation and annihilation over a full oscillation cycle, yielding a finite period-averaged topological charge with reversed sign. For sufficiently large $h_0$ and $h_b$, the re-entrant response disappears and $\langle \chi_Q \rangle \simeq 0$ across the entire bias-field range, and the system evolves into a dynamically polarized state.

The physics of the dynamically re-entrant phase in the dynamic response regime can be understood from the time evolution of the skyrmion number. Figure~\ref{Fig4}(a) depicts $\chi_Q(t)$ over a single field period for different amplitudes $h_0$ at fixed bias $h_b = 0.7$. The total magnetic field $H(t)$ is illustrated in Fig.~\ref{Fig4}(b). For small amplitudes ($h_0 = 0.5$), $\chi_Q(t)$ exhibits only a weak time dependence and remains negative. Increasing $h_0$ induces pronounced oscillations and sign changes in $\chi_Q(t)$, indicating that the external field drives the system between distinct topological states. This is particularly evident for $h_0 = 1.0$ and $1.5$, where significant numbers of skyrmions are dynamically created and annihilated during the field cycle. For $h_0 = 1.5$, corresponding to the yellow region of positive period-averaged skyrmion numbers in Fig.~\ref{Fig3}(b), the positive contribution to $\chi_Q$ that is generated during the first half-period exceeds the negative contribution from the second half-period. This asymmetry in the topological response between successive half-periods arises from the superposition of the oscillatory and bias fields: during the first half-period, the negative oscillatory field dominates over the positive bias field, favoring formation of positive skyrmions, while in the second half-period the fields align driving the system toward a ferromagnetic state. The representative real-space spin textures in Fig.~\ref{Fig4}(c) support this scenario. Spin configurations sampled during the first and second half-periods exhibit positive and negative skyrmions, respectively, followed by a nearly polarized configuration toward the end of the cycle. 
\begin{figure}[h!]
\begin{center}
\includegraphics[scale=0.5]{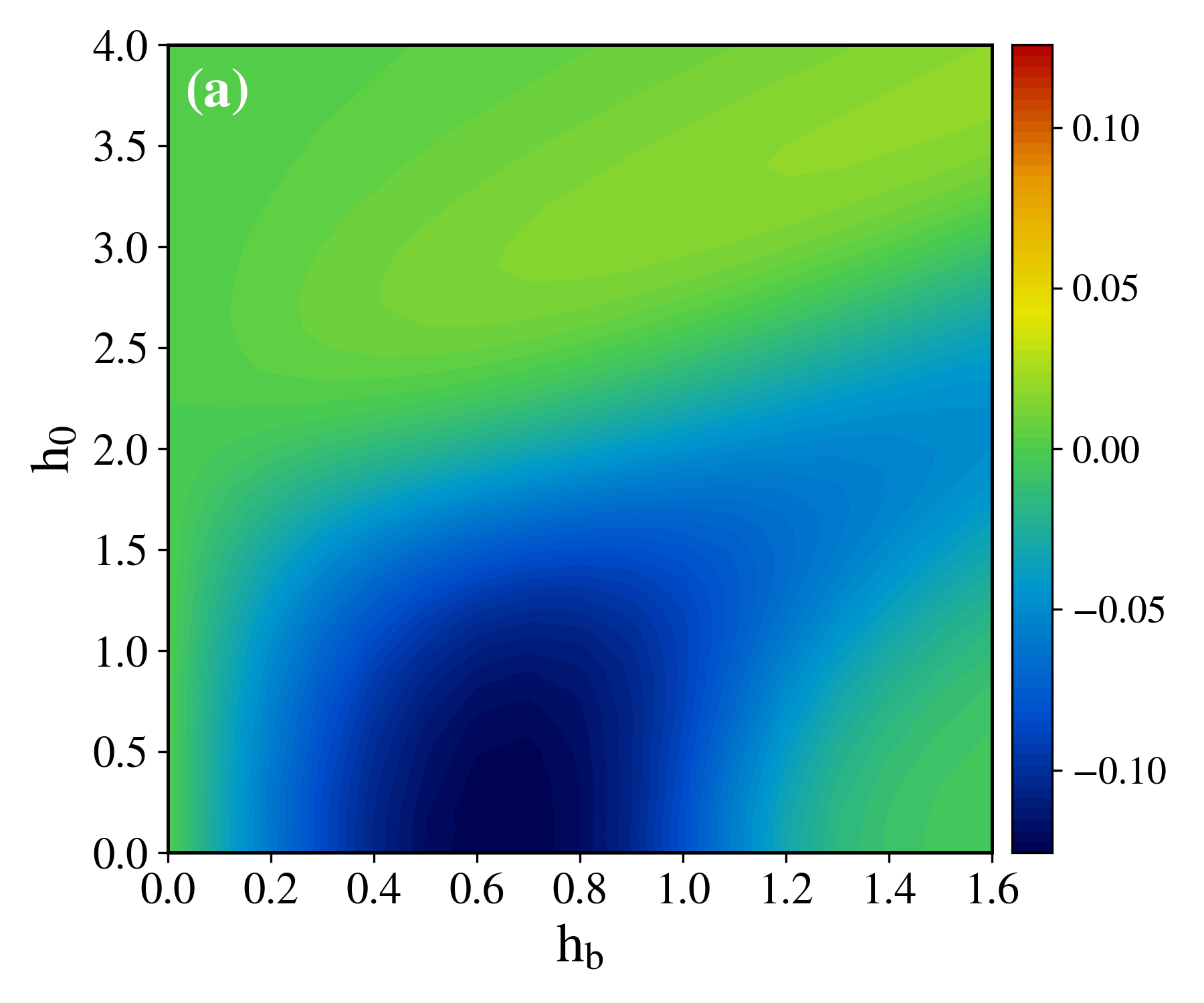}
\includegraphics[scale=0.5]{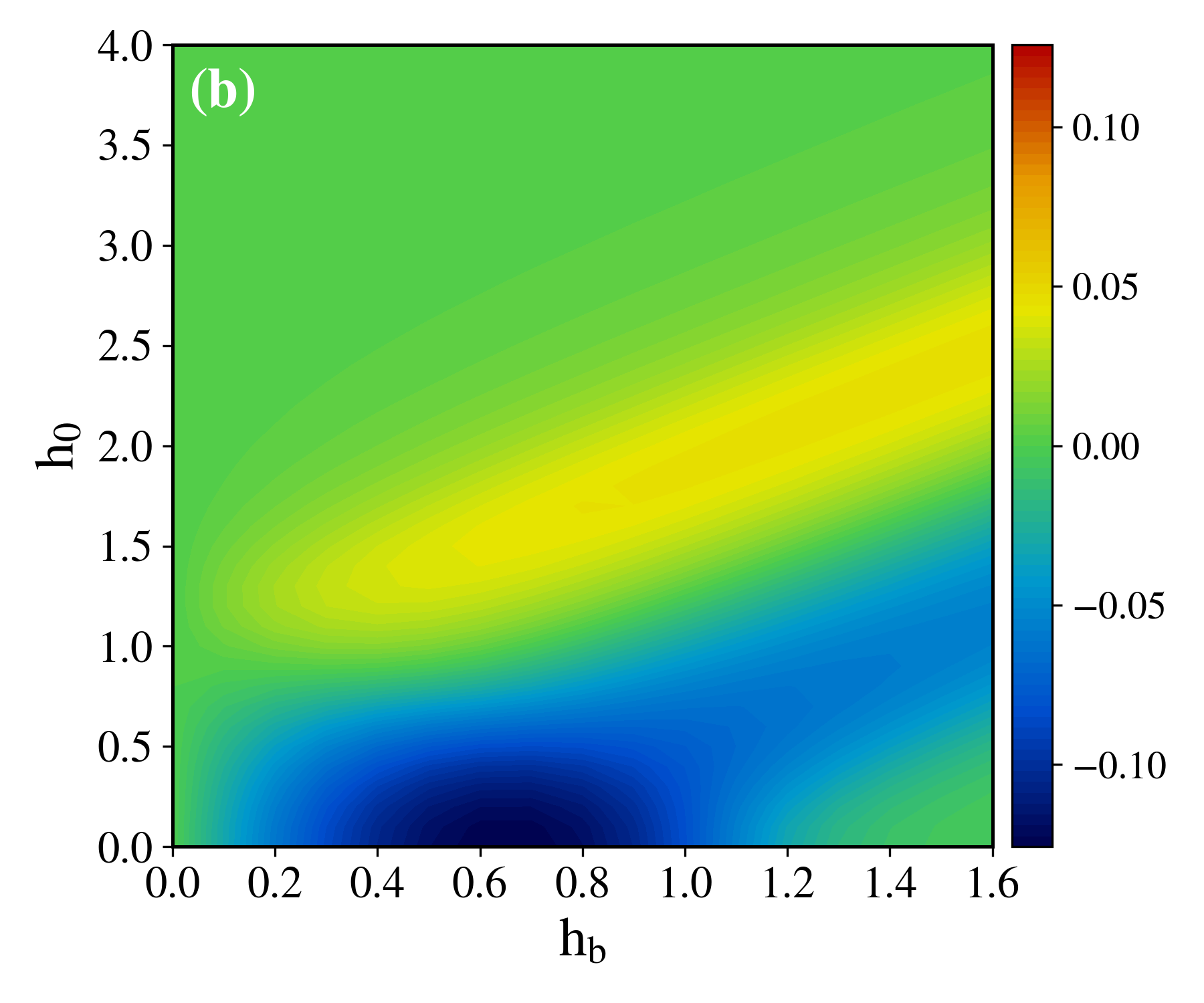}
\caption{Period-averaged skyrmion number $\langle \chi_Q \rangle$ in the $(h_b,h_0)$ plane, computed over a half-period of (a) $t_{1/2}=20$ MCSS, (b) $t_{1/2}=200$ MCSS.}\label{Fig5}
\end{center}
\end{figure}
This interplay between the bias field and the oscillatory drive also explains the inclined, ellipsoidal shape of the dynamically re-entrant skyrmion region, whose slope is approximately unity, indicating that the response is strongest along lines of nearly constant $h_0 - h_b$. Upon further increasing the field amplitude (e.g., $h_0 = 4.0$), the system evolves into a dynamically polarized state, with only small transient deviations in $\chi_Q(t)$ occurring near the field reversals. In equilibrium, skyrmion ensembles may form crystalline or liquid-like states depending on the underlying interactions and thermal fluctuations~\cite{Nishikawa2019, Huang2020, Mitsumoto2022}. In the present simulations, the real-space configurations remain disordered, and no ordered skyrmion lattice emerges within the parameter regime considered. This is consistent with the periodically driven nature of the system, where the dynamic excitation transfers energy periodically to the spin system, thereby making a skyrmion liquid state not only consistent with our observations but also significantly more likely to be realized. Nevertheless, we note that at suitable temperatures and low driving frequencies, transitions into skyrmion lattice states may also be possible.

Finally, we examine the effect of the field period, yet another control parameter of the oscillating magnetic field. Figure~\ref{Fig5} displays the period-averaged skyrmion number $\langle \chi_Q \rangle$ in the $(h_b,h_0)$ plane for different values of the half-period $t_{1/2}$. In the dynamic response regime, corresponding to long half-periods ($t_{1/2}=100$ and $200$ MCSS; the former already shown in Fig.~\ref{Fig3}(b)), a dynamically re-entrant region with positive topological charge (yellow) is clearly observed and becomes more pronounced as $t_{1/2}$ increases. This indicates that the system can respond effectively to the external field during each half-cycle. For shorter periods ($t_{1/2}=20$ MCSS), shown in Fig.~\ref{Fig5}(a), the re-entrant region is strongly suppressed. In this fast-driving regime, the spins cannot follow the rapid field reversals, weakening the asymmetric creation and annihilation of skyrmions within a single cycle. Consistent with this picture, time-series analysis of the skyrmion number at $t_{1/2}=20$ MCSS (see Fig.~S1 in~\cite{supplemental}) shows that the skyrmion dynamics exhibits a reduced ability to follow the oscillating magnetic field relative to the longer half-periods. We note that the dynamically re-entrant skyrmion response also persists at lower temperatures; see Fig.~S2 in~\cite{supplemental}.

In summary, we have investigated the nonequilibrium dynamics of magnetic skyrmions in a two-dimensional ferromagnet driven by periodically oscillating magnetic fields using large-scale Monte Carlo simulations. By analyzing the period-averaged skyrmion number, we identify a dynamically re-entrant topological response in the $(h_b,h_0)$ parameter space that has no equilibrium counterpart. This response, characterized by the suppression and subsequent reemergence of skyrmion order with reversed topological charge, is governed by the field amplitude $h_0$ and exists only in the dynamic response regime, disappearing at higher driving frequencies. Importantly, the dynamically re-entrant behavior represents an intrinsic nonequilibrium response that emerges under an oscillating magnetic field, rather than a transient relaxation effect. In particular, we compare our results with single-quench simulations and discuss the corresponding differences in detail in Ref.~\cite{supplemental}. This behavior remains robust with respect to both the Dzyaloshinskii--Moriya interactions symmetry and the choice of update algorithm and is a characteristic feature of topological magnetic systems under periodic magnetic fields. Our results highlight the richness of nonequilibrium topological phenomena induced by time-dependent fields and establish oscillating magnetic fields as a powerful control parameter for the creation, annihilation, and topological reversal of skyrmion lattice states.

\begin{acknowledgments}
The numerical calculations reported in this paper were performed at T\"{U}B\.{I}TAK ULAKBIM (Turkish agency), High Performance and Grid Computing Center (TRUBA Resources). The work of A.V. and N.G.F. was supported by the  Engineering and Physical Sciences Research Council (grant EP/X026116/1 is acknowledged). Work at nanoGUNE was supported by the Spanish Ministry of Science and Innovation under the Maria de Maeztu Units of Excellence Program (Grant No. CEX2020-001038-M) and Project No. PID2024-155776NB-I00 (ULTRAMAN).
\end{acknowledgments}

%%%%%%%%%%%%%%%%%%%%%%%%%%%%%%%%%%%%%%%%%%%%%%%%%%%%%%%%%%%%%%%%%%%%%%
%%%%%%%%%%%%%%%%%%%%%%%%  SUPPLEMENTAL MATERIAL  %%%%%%%%%%%%%%%%%%%%%
%%%%%%%%%%%%%%%%%%%%%%%%%%%%%%%%%%%%%%%%%%%%%%%%%%%%%%%%%%%%%%%%%%%%%%
\clearpage
\onecolumngrid
\begin{center}
{\large\bfseries Supplemental Material for ``Dynamically Reentrant Skyrmion Phase in Oscillating Magnetic Fields''}\\[4mm]
Z. Demir Vatansever,$^{1}$ E. Vatansever,$^{1}$ A. Vasilopoulos,$^{2}$ N. G. Fytas,$^{2}$ and A. Berger$^{3}$\\[2mm]
{\small\itshape
$^{1}$Department of Physics, Dokuz Eyl\"{u}l University, TR-35160, Izmir-Turkey\\
$^{2}$School of Mathematics, Statistics and Actuarial Science, University of Essex, Colchester CO4 3SQ, United Kingdom\\
$^{3}$CIC nanoGUNE BRTA, E-20018 Donostia-San Sebastian, Spain}
\end{center}
\vspace{4mm}
\twocolumngrid

\setcounter{figure}{0}
\setcounter{table}{0}
\setcounter{equation}{0}
\setcounter{section}{0}
\renewcommand{\thefigure}{S\arabic{figure}}
\renewcommand{\thetable}{S\arabic{table}}
\renewcommand{\theequation}{S\arabic{equation}}
\renewcommand{\thesection}{S\Roman{section}}
\section{Lattice discretization of the topological charge}

\begin{figure}[h!]
\begin{center}
\includegraphics[scale=0.4]{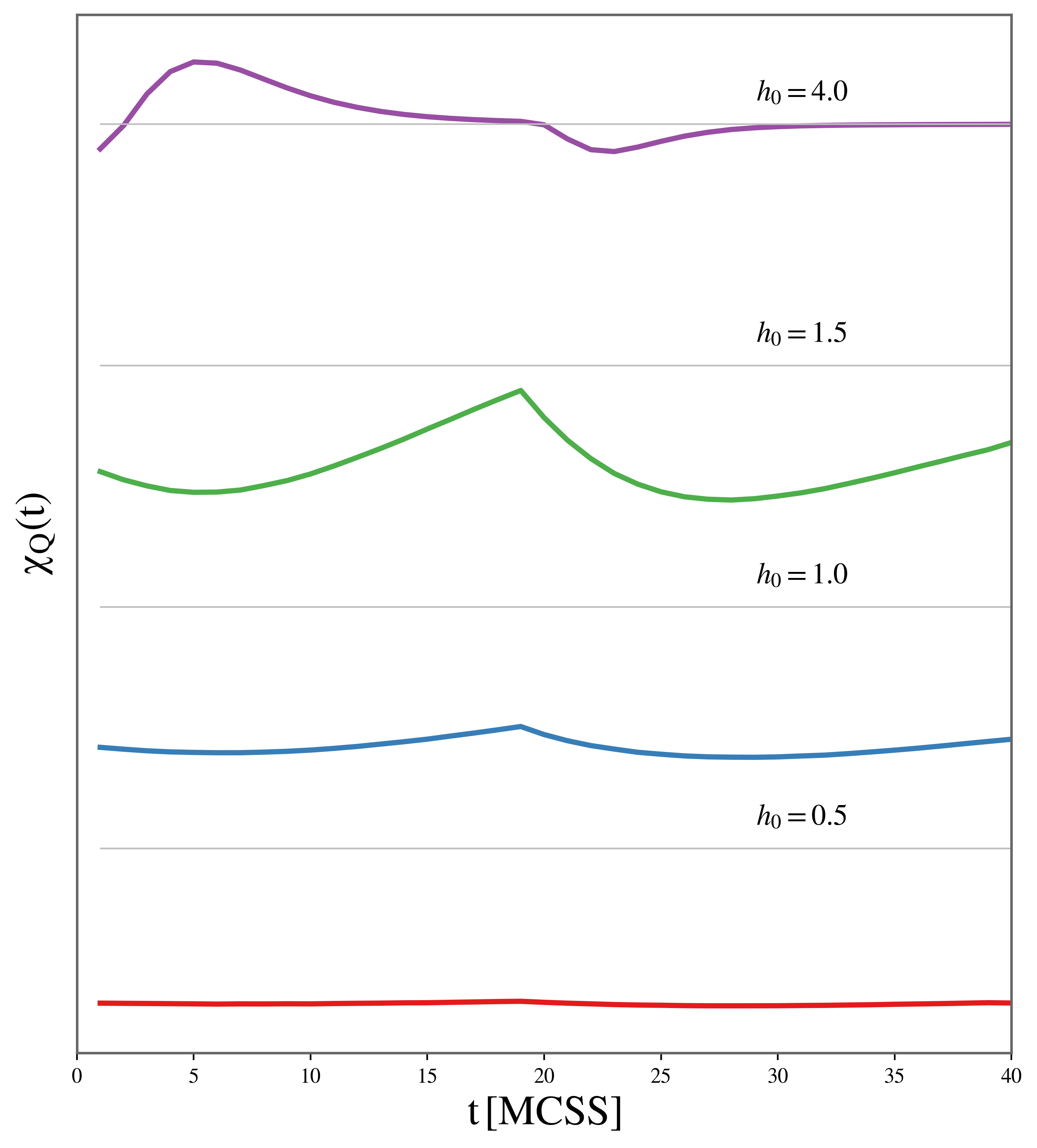}
\caption{Time series of the skyrmion number $\chi_Q(t)$ for several values of $h_0$ at a half-period $t_{1/2}=20$ MCSS. The gray horizontal line represents $\chi_Q = 0$.} \label{Fig1S}
\end{center}
\end{figure}

The topological charge $\chi_Q$ is evaluated on the triangular lattice using a discretization based on elementary triangular plaquettes. Each lattice site $i$ is associated with two oriented triangles, labeled $(i,1,2)$ and $(i,3,4)$, as illustrated in Fig.~1 of the main text, ensuring a complete and non-overlapping tiling of the lattice.
The total chirality, $\chi_L$, is defined as
\begin{equation}
\chi_L = \frac{1}{8\pi} \Bigg\langle \sum_{\mathbf{r}_i}
\chi_{L,i}^{(12)} + \chi_{L,i}^{(34)} \Bigg\rangle .
\end{equation}
Here, $\chi_{L,i}^{(ab)} = \mathbf{S}_{i} \cdot (\mathbf{S}_{a} \times \mathbf{S}_{b})$ denotes the local scalar chirality associated with the oriented triangle formed by the lattice sites $(i,a,b)$. In this notation, $i$ labels the central lattice site, while $1\sim 4$ denote its nearest neighbors involved in the construction of the two elementary triangles. 
The discretized skyrmion number $\chi_Q$, a topological quantity that characterizes the winding of a spin configuration in a two-dimensional system, is defined as
\begin{equation}
\chi_Q = \frac{1}{4\pi} \Bigg\langle \sum_{\mathbf{r}_i}
A_{i}^{(12)} \, \mathrm{sgn}\!\left[\chi_{L,i}^{(12)}\right]
+ A_{i}^{(34)} \, \mathrm{sgn}\!\left[\chi_{L,i}^{(34)}\right] \Bigg\rangle .
\end{equation}
Here, $A_{i}^{(ab)} = \|\mathbf{S}_{a} - \mathbf{S}_{i} \times \mathbf{S}_{b} - \mathbf{S}_{i}\|/2$ is the local area of the surface spanned by three spins on each elementary triangle. In Fig.~\ref{Fig1S} we depict some typical time series of the skyrmion number $\chi_Q(t)$ for several values of $h_0$ at a half-period $t_{1/2} = 20$ Monte Carlo steps per site (MCSS).

\section{Metropolis update protocol}

In our Monte Carlo simulations, the spin configurations are updated using the standard Metropolis algorithm. During each MCSS, lattice sites are visited randomly and a trial update is proposed for each spin. The trial spin orientations are generated randomly over the unit sphere using the Marsaglia procedure~\cite{Marsaglia1972}, which provides an efficient method for sampling uniformly distributed random unit vectors. The resulting transformation yields a unit vector that is uniformly distributed over the surface of the sphere. The newly generated unit vector is then used as the trial spin orientation. The trial move is accepted with the standard Metropolis probability
\[
P_{\mathrm{acc}}=\min\left[1,\exp(-\beta \Delta E)\right],
\]
where $\Delta E$ is the energy difference between the trial and current configurations and $\beta=1/(k_B T)$.

To test the reliability of our results with respect to the update procedure, we have performed additional simulations in the re-entrant region of the phase diagram using the heat-bath algorithm~\cite{Nishikawa2019}. The dynamically re-entrant behavior is also observed with the heat-bath update scheme, which indicates that the main physical picture reported in the manuscript does not depend on the specific choice of update algorithm. We note, however, that in Monte Carlo studies of dynamic phase transitions, time-dependent quantities such as the metastable lifetime and the critical half-period may depend on the update rule~\cite{Buendia3, Tauscher2014}. Consequently, while the heat-bath simulations confirm the existence of the re-entrant behavior, quantitative details such as the precise location of the phase boundaries may show small algorithm-dependent shifts.

\section{Low-temperature results of period-averaged dynamic skyrmion numbers}

\begin{figure}[h!]
\begin{center}
\includegraphics[scale=0.4]{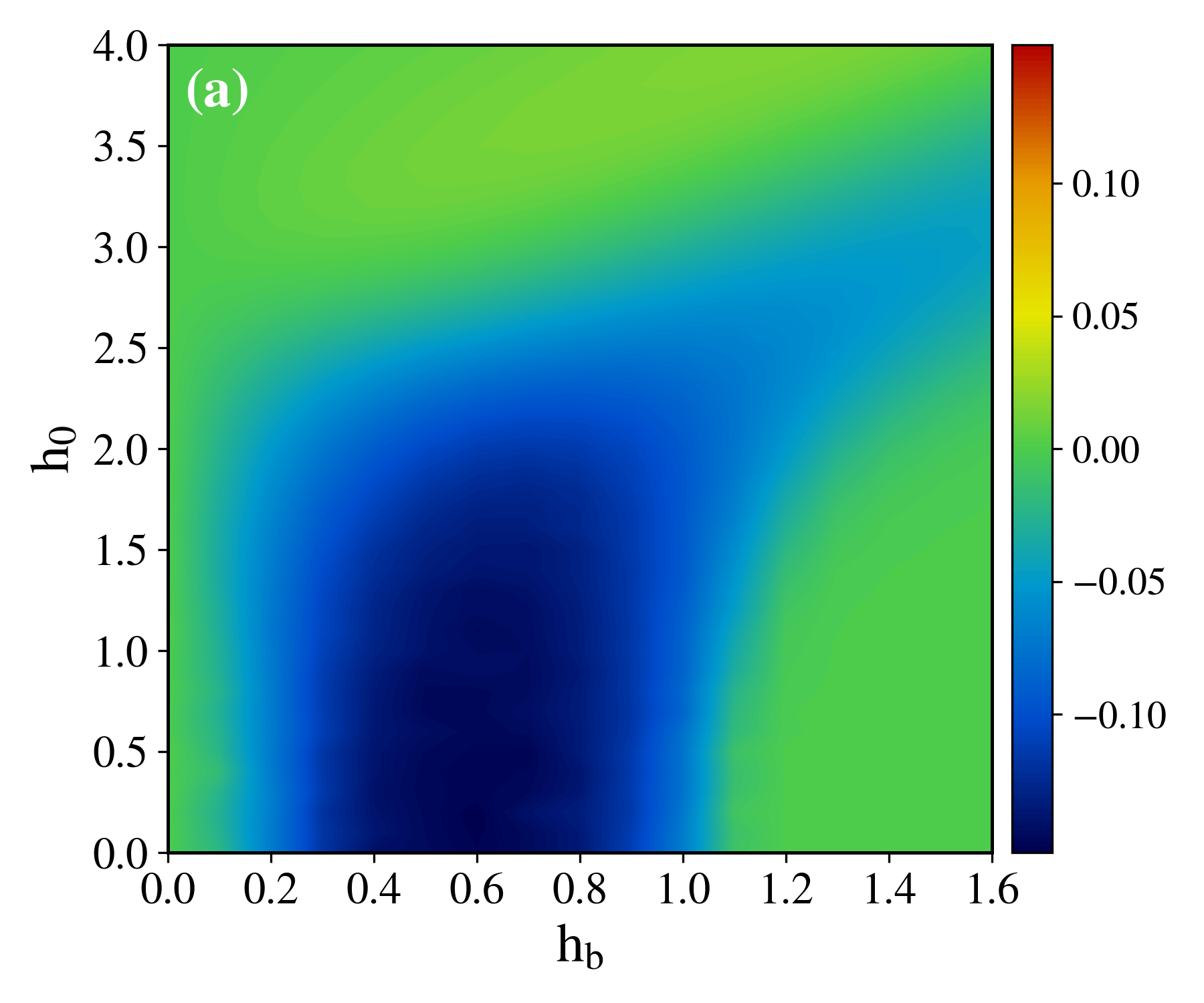}
\includegraphics[scale=0.4]{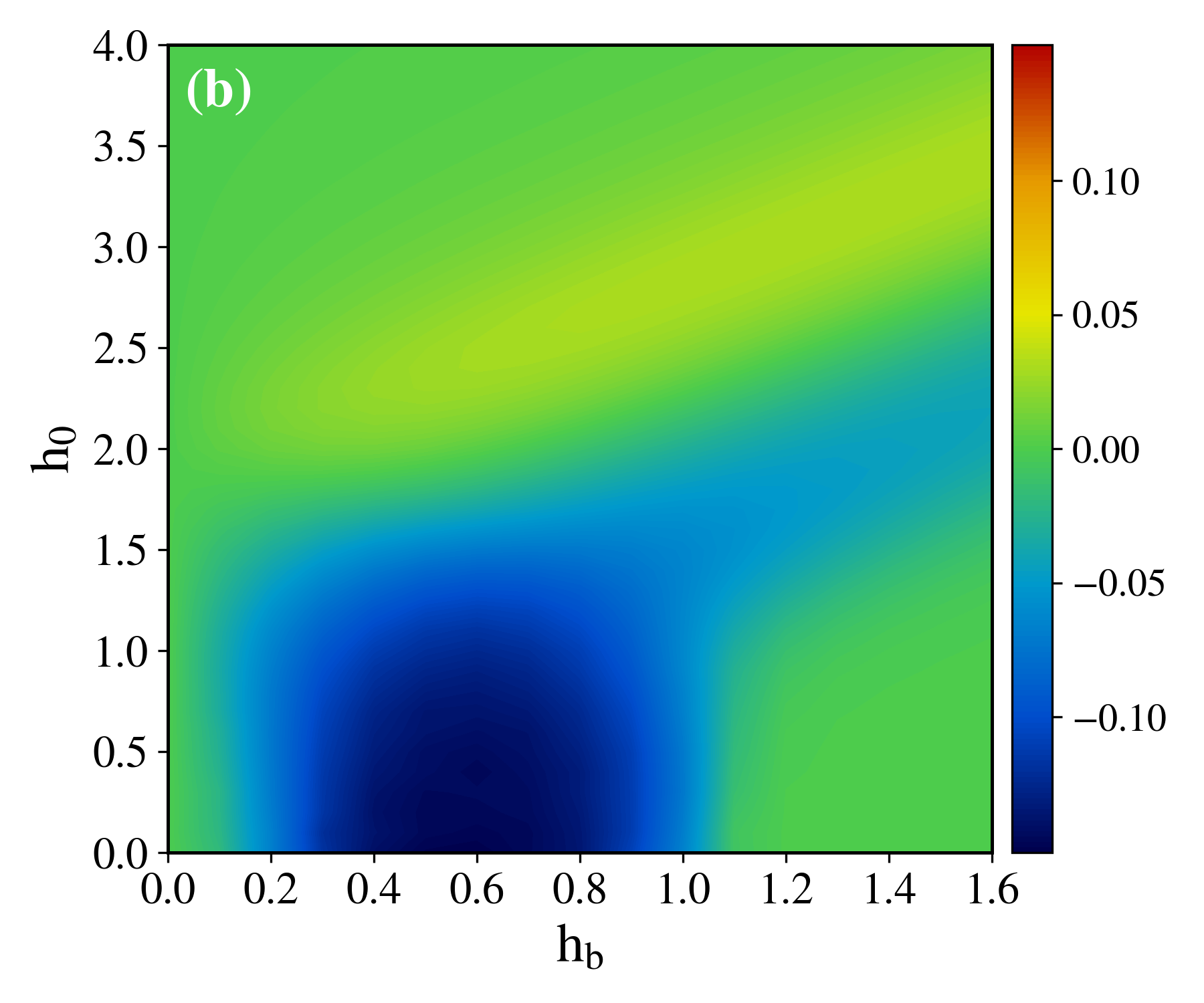}
\includegraphics[scale=0.4]{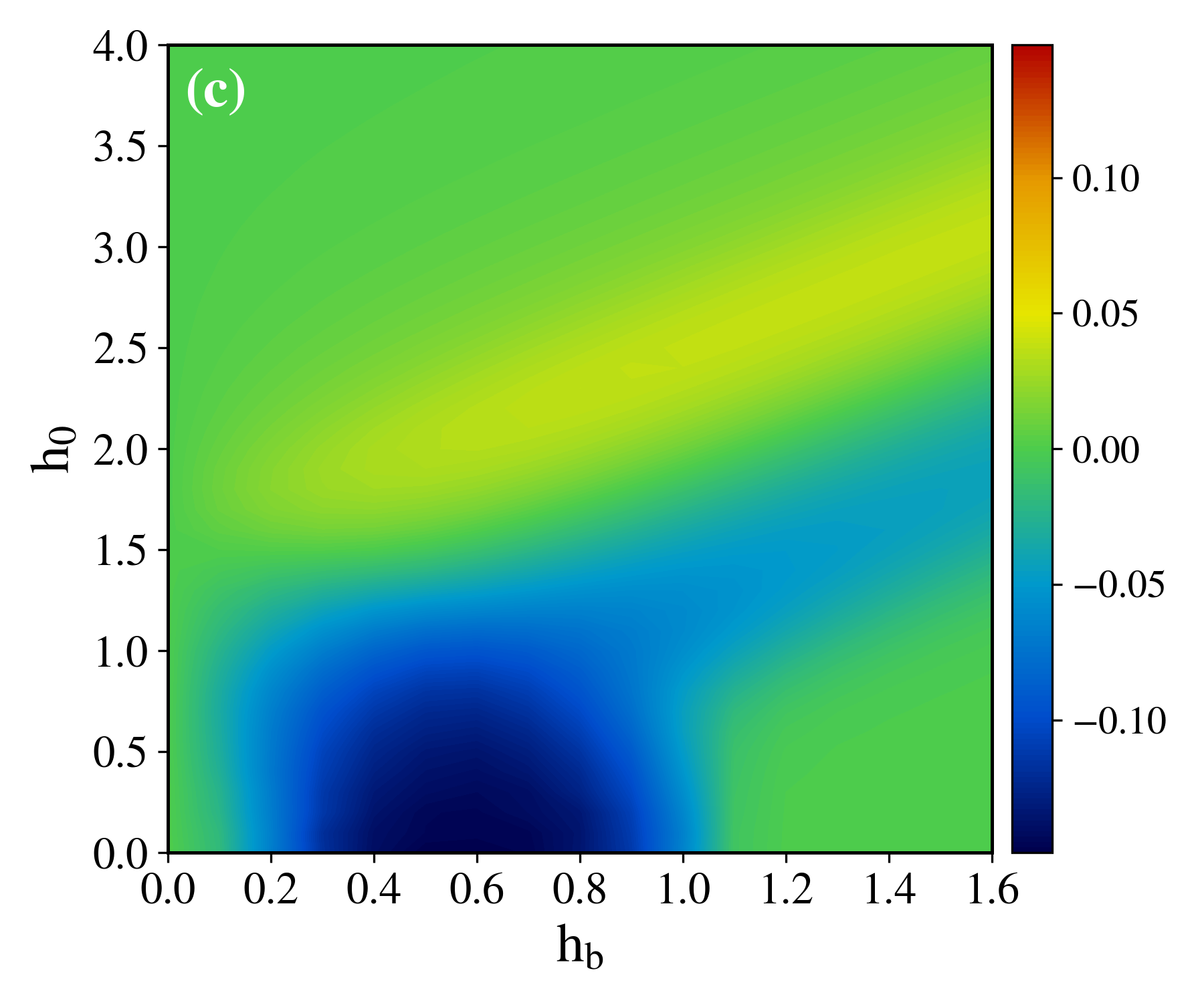}
\caption{Period-averaged skyrmion number $\langle \chi_Q \rangle$ (per $10^{2}$ unit cells) in the $(h_b,h_0)$ plane, calculated over a half-period of (a) $t_{1/2}=20$ MCSS, (b) $t_{1/2}=100$ MCSS, and (c) $t_{1/2}=200$ MCSS at $T=0.5$. The overall phase diagrams obtained for $T=1.0$ (shown in the main text)  remain unchanged, while regions with negative $\langle \chi_Q \rangle$ expand and the positive regions shift to higher field amplitudes for $T=0.5$. This behavior reflects the reduced thermal fluctuations at lower temperatures, which enhance the stability of topological textures.}\label{Fig2S}
\end{center}
\end{figure}

In Fig.~\ref{Fig2S} we present some additional low-temperature results of period-averaged dynamic skyrmion numbers in the ($h_b$, $h_0$) plane.

\section{Effect of Dzyaloshinskii--Moriya Interactions Symmetry on the Dynamically Re-entrant Behavior}

\begin{figure}[h!]
\centering
   \includegraphics[scale=0.5]{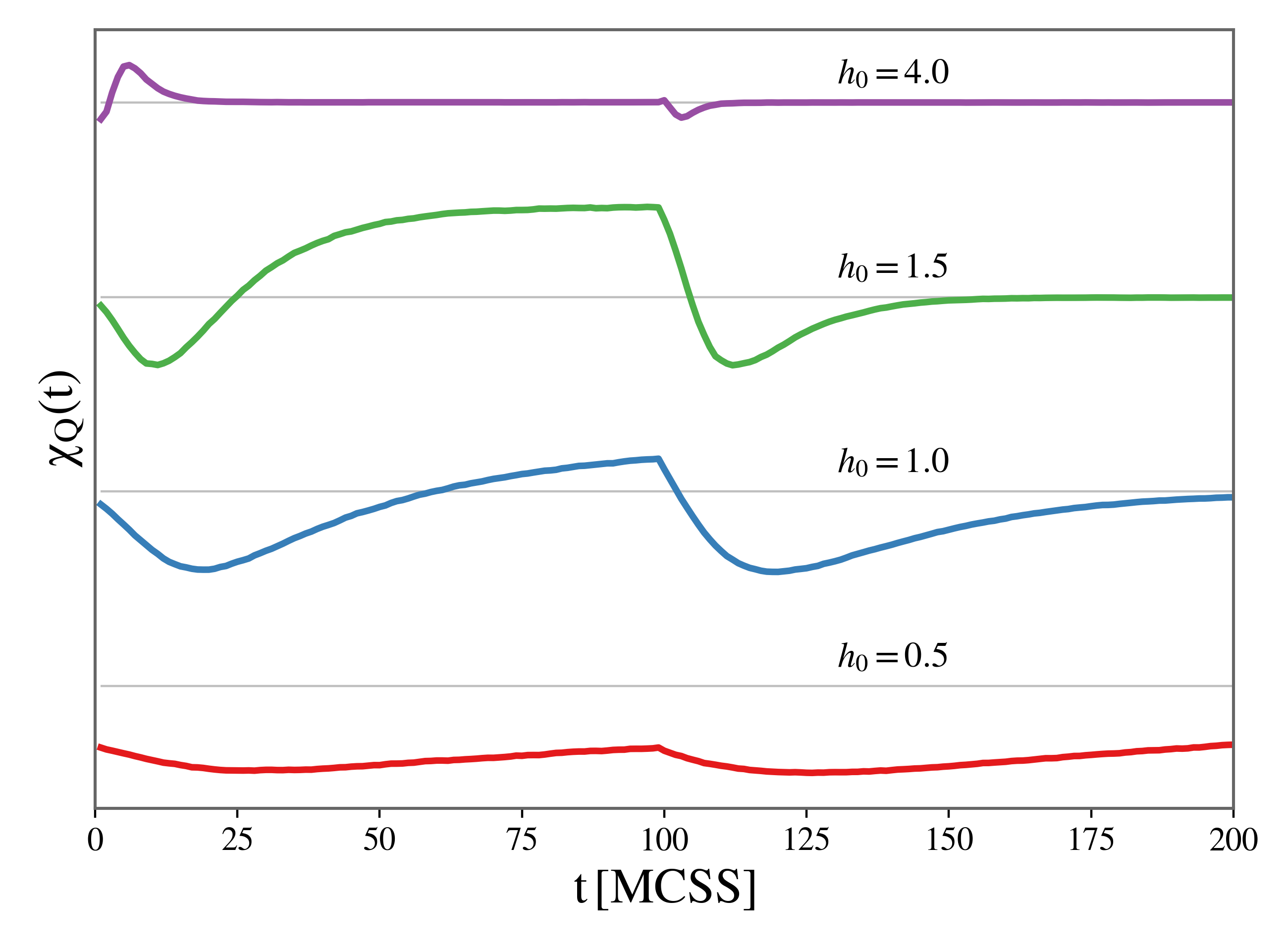}
\caption{Time evolution of the skyrmion number $\chi_Q(t)$ for the Bloch-type DMI configuration for a half-period of $t_{1/2}=100$ MCSS. The curves correspond to $h_b = 0.7$ and different field amplitudes $h_0 = 0.5$, $1.0$, $1.5$, and $4.0$. The gray lines represent $\chi_Q=0$.}
\label{FigR3}
\end{figure}

We also test the effect of Dzyaloshinskii--Moriya interactions (DMI) symmetry by replacing the Néel-type DMI with the Bloch-type form, keeping all other parameters unchanged. Figure~\ref{FigR3} shows the time evolution of the skyrmion number for several $h_0$ values. The dynamically re-entrant behavior is also observed for Bloch-type DMI as seen at $h_0 = 1.5$, and is therefore not restricted to a specific DMI symmetry.

\begin{figure}[hbt!]
\centering
\includegraphics[scale=0.46]{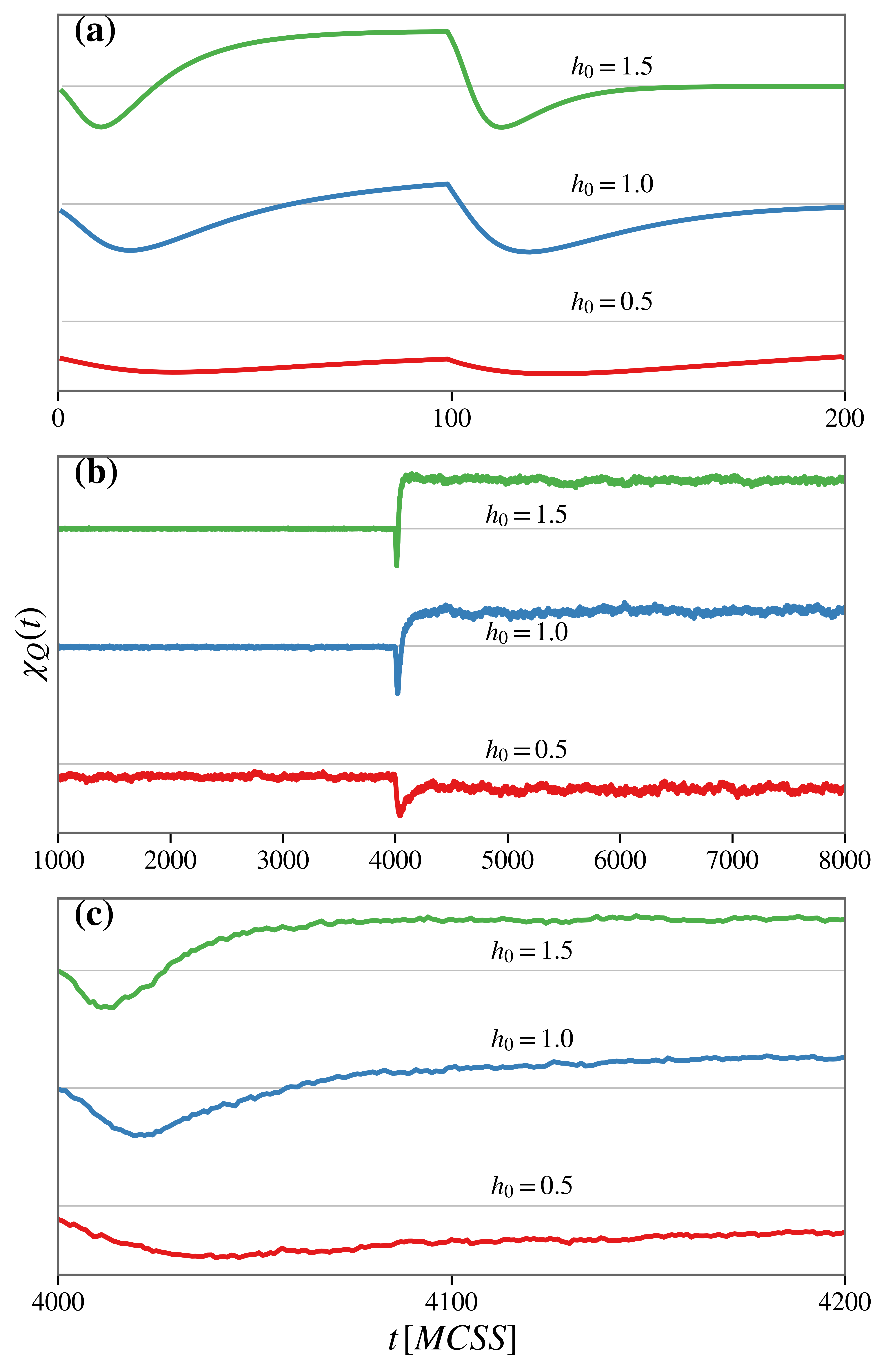}
\caption{Time evolution of the skyrmion number $\chi_Q(t)$ for $h_b=0.7$ and different values of the field amplitude $h_0$. (a) Periodically driven system under an oscillating magnetic field for a half period of $t_{1/2}=100$ MCSS. (b) Dynamics obtained using the single-quench protocol, where the system is first equilibrated at $h_b+h_0$, and the magnetic field is then suddenly switched to $h_b-h_0$ at $t=4000$ MCSS. (c) Zoomed-in view of panel (b) in the time range $4000-4200$ MCSS. The gray lines represent $\chi_Q=0$.}
\label{FigS3}
\end{figure}

\section{Comparison between Periodically Oscillating Field and Single-Quench Dynamics}

In this section, we compare the dynamics under a periodically oscillating magnetic field with those obtained from a single-quench protocol. We analyze the dynamics obtained from a single-quench protocol, in which the system is first equilibrated at the high field $h_b + h_0$ and the magnetic field is then suddenly switched to $h_b - h_0$.

\begin{figure}[hbt!]
\centering
   \includegraphics[scale=0.5]{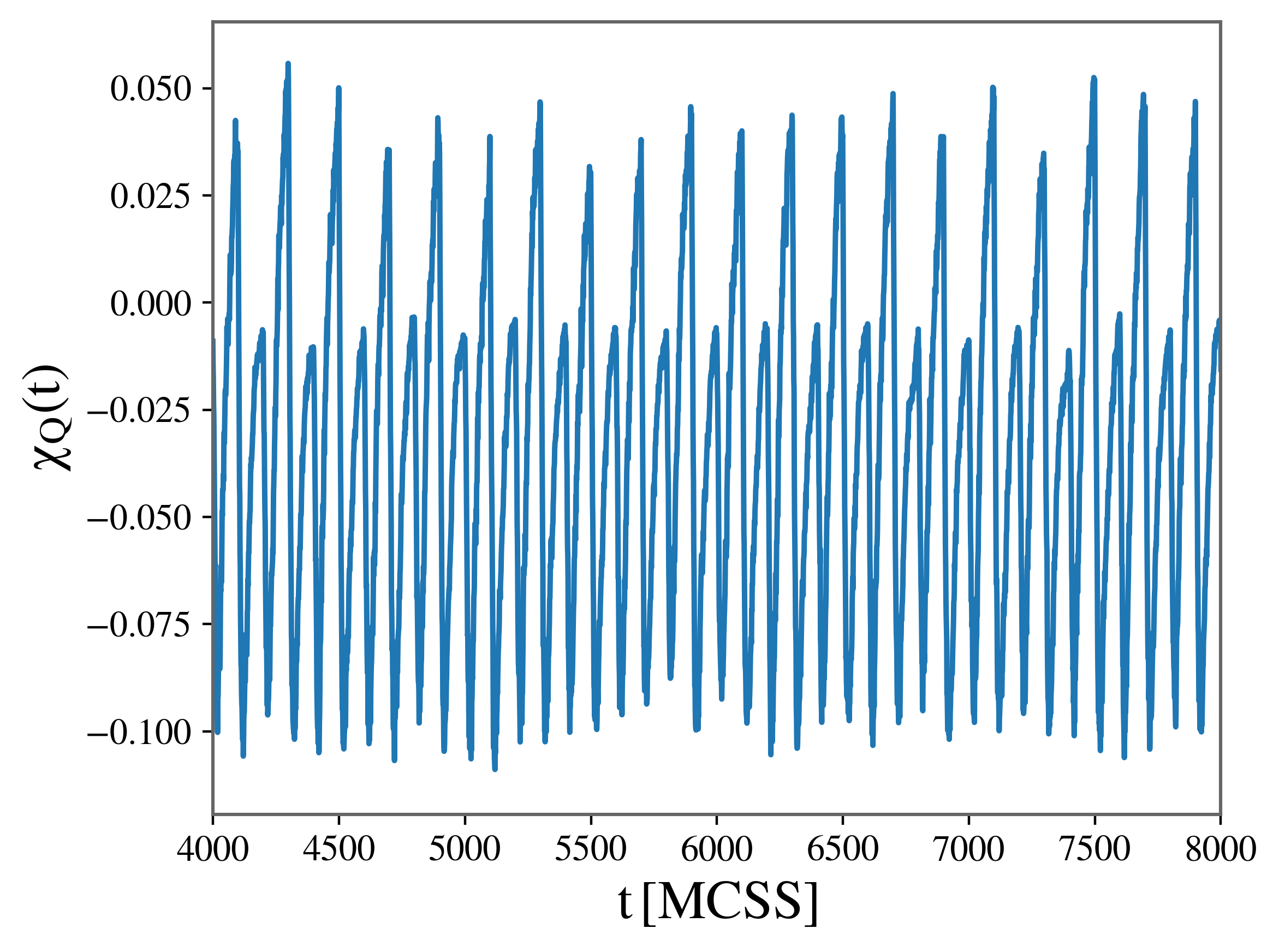}
\caption{Long time evolution of $\chi_Q(t)$ under a periodically oscillating magnetic field for $h_b=0.7$, $h_0=1.0$, and $t_{1/2}=100$ MCSS.}\label{FigR2}
\end{figure}

Figure~\ref{FigS3}(a) shows the dynamics under a periodic field, while Fig.~\ref{FigS3}(b,c) correspond to the single-quench protocol, where the system relaxes from a strongly polarized state to a stationary state. Comparing Fig.~\ref{FigS3}(a) with Fig.~\ref{FigS3}(c), one observes that the transient dynamics in the single-quench case (around $t=4000$–$4100$ MCSS) resemble those of the first half-cycle under a periodic field. However, this behavior remains transient and does not develop into a persistent cycle-to-cycle sequence of annihilation and creation processes. In the single-quench protocol, a time-dependent evolution of the skyrmion state can occur, and even an inversion may be observed, for example, at larger field amplitudes such as $h_0 = 1.0$ or $h_0 = 1.5$, in the transient regime. In contrast, under a periodic magnetic field, this behavior emerges within each field cycle and is repeated, leading to persistent switching between skyrmion states. Moreover, the single-quench dynamics may be misleading. Although the relaxation curves for $h_0 = 1.0$ and $h_0 = 1.5$ appear qualitatively very similar, the corresponding time-averaged dynamic order parameter under oscillatory excitation is actually inverted in between these two cases. In addition, its detectability is crucially different if one were to search for experimental verifications of the here-reported dynamically re-entrant behavior. Observations of the transient inversion in a single-quench protocol would require extremely high time resolution, likely to be on the picosecond scale, as the actual trajectory would have to be measured in real time or by means of pump-probe type experiments. In contrast, periodic driving makes the dynamically re-entrant behavior observable by means of time-averaged quantities, which are obviously much more accessible in actual measurements.

Figure~\ref{FigR2} shows the long-time evolution of the skyrmion number under a periodically oscillating magnetic field. As shown in Fig.~\ref{FigS3}(a), the time evolution over one period is obtained by averaging over many field cycles in the long-time regime. This is fundamentally different from averaging over many independent realizations of the single-quench protocol, which only reduces statistical fluctuations across realizations. Under a periodic field, fluctuations arise within a single realization and are synchronized with the field, yielding similar dynamical patterns in successive cycles. In the single-quench protocol, run-to-run fluctuations originate from differences between independent realizations and do not represent a repeated dynamical process.


\begin{thebibliography}{99}
\bibitem{ParExp1} Here $k_{\rm B}=1$ is a choice of units, and fixing $J=1$ sets the exchange Dzyaloshinskii-Moriya interaction energy scale.
\end{thebibliography}
\end{document}